\documentclass[aps,pra,twocolumn,showpacs,preprintnumbers,amsmath,amssymb]{revtex4-1}
\usepackage{graphicx} 
\usepackage[usenames]{color}

\begin{document}
\preprint{INT-PUB-15-008}
\title{Gauge-invariant implementation of the Abelian Higgs model on optical lattices}
\author{A. Bazavov$^{1,2}$}
\author{Y. Meurice$^1$}
\author{S.-W. Tsai$^2$}
\author{J. Unmuth-Yockey$^1$}
\author{Jin Zhang$^2$}
\affiliation{$^1$ Department of Physics and Astronomy, The University of Iowa, Iowa City, Iowa 52242, USA }
\affiliation{$^2$ Department of Physics and Astronomy, University of California, Riverside, CA 92521, USA}

\def\lt{\lambda ^t}
\def\note{note}
\def\beq{\begin{equation}}
\def\enq{\end{equation}}

\date{\today}
\begin{abstract}
We present a gauge-invariant effective action for the Abelian Higgs model (scalar electrodynamics) with a chemical potential $\mu$ on a 1+1 dimensional  lattice.
This formulation provides an expansion in the hopping parameter $\kappa$ which we test with Monte Carlo simulations
for a broad range of the inverse gauge coupling $\beta_{pl}$ and small values of the scalar self-coupling $\lambda$. 
In the opposite  limit of infinitely large $\lambda$, the partition function can be written as a traced product of local tensors which allows us to write exact blocking formulas. 
Their numerical implementation requires truncations but there  
is no sign problem for arbitrary values of $\mu$. 
We show that 
the time continuum limit of the blocked transfer matrix can be obtained numerically and, in the limit of infinite $\beta_{pl}$ and with a spin-1 truncation, the small volume energy spectrum is identical to the low energy spectrum of a two-species Bose-Hubbard model in the limit of large onsite repulsion. We extend this procedure for finite $\beta_{pl}$ and derive a spin-1 approximation of the Hamiltonian. It involves new terms corresponding to transitions among the two species in the Bose-Hubbard model. We propose an optical lattice implementation involving a ladder structure. 
\end{abstract}

\pacs{05.10.Cc,05.50.+q,11.10.Hi,11.15.Ha,64.60.De,75.10.Hk }
\maketitle

\section{Introduction}
\label{sec:intro}

The lattice formulation of quantum chromodynamics has provided successful treatments of nonperturbative  problems involving strongly interacting particles in the 
context of particle and nuclear physics. However, numerical computations 
at finite density and real time have remained a major challenge.  For this reason, there has been a lot of interest in the possibility of building quantum simulators for lattice gauge theory (LGT) using optical lattices (for recent reviews see Refs.  \cite{uwreview,Zohar:2015hwa}).

When physical degrees of freedom - individual cold atoms or condensates trapped  in the optical lattice - play the role of the gauge fields, one needs to make sure that the physical observables are gauge invariant. One way to achieve this 
goal is to impose Gauss's law at least in some approximation. This question is discussed  in the context of Abelian gauge theories in 
Refs. \cite{PhysRevA.83.033625,Zohar:2011cw,Zohar:2012ay,Tagliacozzo:2012vg,Kasamatsu:2012im,uwmask,Kuno:2014npa}.
Another approach  that has been advocated \cite{Meurice:2012wk,Liu:2012dz} is to work directly with a gauge invariant formulation. 

A simple example 
is the correspondence between the Fermi-Hubbard model and a $SU(2)$ gauge theory with one fermion \cite{PhysRevB.38.2926}. At strong gauge coupling and small hopping parameter, the quadratic part of the gauge-invariant effective Hamiltonian in one spatial dimension reads
\begin{equation}
H_{eff}\propto \sum_{i}[M_i M_{i+1}+2(B_i^\dagger B_{i+1}+B_{i+1}^\dagger B_{i})] \ ,
\label{eq:mandb}
\end{equation} 
where $M_i$ and $B_i$ are the gauge invariant operators corresponding to mesons and $SU(2)$-baryons. 
A similar effective Hamiltonian can be obtained at second order in degenerate perturbation theory of the 
Fermi-Hubbard model with strong onsite repulsion. 

Effective actions for mesons and baryons can be constructed for $SU(N)$ gauge groups with similar approximations~\cite{Rossi:1984cv}. The computation of the corrections due to the plaquette interactions is rather involved~\cite{deForcrand:2013ufa}.
Various related techniques have been developed to approach chiral symmetry breaking at strong coupling 
\cite{Tomboulis:2012nr,Christensen:2015axa} and can accommodate a chemical potential \cite{Tomboulis:2013ppa}. Despite this progress,  
testing the validity of the approximations with numerical methods has remained challenging which suggests to consider simpler examples. 

For the Bose-Hubbard model with one species of particle,  there is a remarkable level of {\it quantitative} agreement \cite{trot}   between state of the art quantum Monte Carlo (MC) 
calculations and their experimental optical lattice implementations. 
It would be highly desirable to provide a similar proof of principle for a simple LGT model where calculations at finite density and real time are possible. 

In this article, we explain how to establish an approximate {\it quantitative} correspondence between the well-known  Abelian Higgs model (scalar electrodynamics) \cite{RevModPhys.52.453,PhysRevD.19.1882} with a chemical potential $\mu$ on a 1+1 dimensional  lattice and specific many-body theories that can in principle be realized experimentally on optical lattices.  The model describes a charged complex scalar field and an Abelian  gauge field. Its formulation is reviewed in Sec. \ref{sec:abh}. The model can be studied numerically by 
various numerical methods in 
various regions of the parameter space and a first important step is to connect the standard isotropic space-time calculations to the Hamiltonian 
approach obtained by taking the time continuum limit first and then connecting this Hamiltonian with the Hamiltonian of cold atoms trapped in optical lattices. Our first goal  is to achieve some calibration in the gapped, confining phase corresponding to the Mott insulator phase of the cold atom system. 

The exact form of the gauge-invariant effective action for the mesons is derived in Sec. \ref{sec:eff} by integrating over the gauge fields. 
The absence of fermions permits 
inexpensive  MC simulations at Euclidean time and zero chemical potential. This allows us to test  the expansion 
of the effective action in the hopping parameter 
for a broad range of the inverse gauge coupling $\beta_{pl}$ and small values of the scalar self-coupling $\lambda$. Good agreement with MC is shown in Sec. \ref{sec:MC}.

In Sec. \ref{sec:lambda} we consider the opposite  limit of arbitrarily large $\lambda$ where the amplitude of the scalar field is frozen to unity. In this limit, all the remaining variables involved in the basic formulation are compact and we can use recently developed Tensor Renormalization Group (TRG) methods \cite{PhysRevB.86.045139,PhysRevD.88.056005,signtrg,PhysRevE.89.013308} to write 
the partition function as a traced product of local tensors with discrete indices. This reformulation 
allows us to write exact blocking formulas which can be used for numerical purposes. 
The practical implementation requires truncations but in contrast to the MC approach there is no sign problem for arbitrary values of $\mu$. We then write the transfer matrix as a product of tensors along a time slice.

The time continuum limit of the blocked transfer matrix can be obtained numerically.  In the limit of infinite $\beta_{pl}$ and with a spin-1 truncation (inspired by gauge magnet or gauge link constructions \cite{Orland:1989st,Chandrasekharan:1996ih}), the small volume energy spectrum is identical to the low energy spectrum of a two-species Bose-Hubbard model in the limit of large onsite repulsion \cite{PhysRevA.90.063603,Bazavov:2014lla}. We calculate numerically the 
energy spectrum in this limit in Sec. \ref{sec:weak}.

In Sec. \ref{sec:time}, we extend this procedure for finite $\beta_{pl}$ and derive a spin-1 approximation of the Hamiltonian. It involves new terms corresponding to transitions among the two species in the Bose-Hubbard model which we discuss in Sec. \ref{sec:bh2}. We then propose an optical lattice implementation involving a ladder structure and discuss further plans to 
obtain a good correspondence between conventional MC calculations and optical lattice measurements.

\section{The Abelian Higgs model}
\label{sec:abh}
In this section, we briefly remind the reader of the action for the Abelian Higgs model
on a 1+1 space-time lattice of size $N_s\times N_\tau$  and introduce the notations used later. We use $x$, $y$ etc.  for space-time vectors, $i$, $j$ etc. for the one-dimensional spatial sites and $\hat{\nu}$= $\hat{s}$ or $\hat{\tau}$ for the unit vectors in 
space and time, respectively. The gauge fields 
$U_{x,\hat{\nu}}=\exp{i A_{x,\hat\nu }}$ are attached to the links. We denote the product of $U$'s around a plaquette $U_{pl,x}$ where $x$ is the lower left corner of the plaquette in space-time coordinates. We use the notation $\beta_{pl}=1/e^2$ for the inverse gauge coupling and $\kappa_s$ ($\kappa_\tau$) for the hopping 
coefficient in the space (time) direction. 
For the potential for the complex scalar field $\phi_x=|\phi_x|\exp(i\theta_x)$, we follow the convention of Ref.~\cite{heitgerphd}. 
The action reads:
\begin{equation}\label{S_u1h}
S = S_g + S_h + S_\lambda,
\end{equation}
where the gauge part is
\begin{equation}
S_g = -\beta_{pl}\sum_x{\rm Re}\left[U_{pl,x}\right],
\end{equation}
the hopping
\begin{eqnarray}
S_h &=& -{\kappa_\tau}\sum_x
\left[{\rm e}^{\mu}\phi_x^\dagger U_{x,\hat\tau}\phi_{x+\hat\tau}+{\rm e}^{-\mu}
\phi_{x+\hat\tau}^\dagger U^\dagger_{x,\hat\tau}\phi_x
\right] \nonumber\\ 
&-&{\kappa_s}\sum_x
\left[\phi_x^\dagger U_{x,\hat{s}}\phi_{x+\hat{s}}+
\phi_{x+\hat{s}}^\dagger U^\dagger_{x,\hat{s}}\phi_x
\right]
\end{eqnarray}
and the self-interaction
\begin{equation}
S_\lambda = \lambda\sum_x\left(\phi_x^\dagger\phi_x-1\right)^2+
\sum_x\phi_x^\dagger\phi_x .
\end{equation}
The partition function can then be written as
\begin{equation}
Z=\int D\phi^\dagger D\phi DU e^{-S} .\end{equation}
In the hopping part of the action $S_h$, we can separate the compact and non-compact variables
\begin{eqnarray}
S_{h}=&-&  2\kappa_\tau |\phi_x||\phi_{x+\hat\tau}| \sum\limits_{x} \cos(\theta_{x+\hat\tau} - \theta_{x}+A_{x,\hat\tau}-i\mu)\cr&-&
2\kappa_s |\phi_x||\phi_{x+\hat{s}}| \sum\limits_{x} \cos(\theta_{x+\hat{s}} - \theta_{x}+A_{x,\hat{s}}) .
\end{eqnarray}
This equation makes clear that the chemical potential is a constant imaginary gauge field in the time direction and that the Nambu-Goldstone fields $\theta_x$ can be eliminated by a gauge transformation 
\begin{equation}
A_{x,\hat\nu}\rightarrow A_{x,\hat\nu}-\theta_{x+\hat\nu} + \theta_{x},
\end{equation}
which leaves the plaquette terms unchanged.

\section{A gauge-invariant effective action }
\label{sec:eff}
As explained in the introduction, unlike other approaches~\cite{PhysRevLett.110.125303, PhysRevLett.110.125304,Tagliacozzo:2012vg,Wiese:2013uua,Zohar:2015hwa,Zohar:2011cw,Zohar:2012ay,Kasamatsu:2012im,Kuno:2014npa} we will not try to implement the gauge field on the optical lattice, but rather try to implement a gauge-invariant effective action obtained by integrating over the gauge fields. 
In this section, we will show that 
this effective action is a function of the composite, gauge-invariant, meson field which we denote  $M_x\equiv\phi_x^\dagger \phi_x$ in order to emphasize the analogy with Eq. (\ref{eq:mandb}). In other words, 
\begin{equation}
Z=\int D\phi^\dagger D\phi DU e^{-S}=\int DMe^{-S_{eff}({M})-S_\lambda(M)}.
\end{equation}
For this purpose, we use the Fourier expansion of the Boltzmann weights in terms of the modified Bessel functions $I_n$, for instance, 
\begin{eqnarray}
&\ &\exp[2\kappa_\tau |\phi_x||\phi_{x+\hat\tau}|  \cos(\theta_{x+\hat\tau} - \theta_{x}+A_{x,\hat\tau}-i\mu)]\\
&=&\sum_{n=-\infty} ^{\infty}I_n(2\kappa_\tau |\phi_x||\phi_{x+\hat\tau}|)\exp[in(\theta_{x+\hat\tau} - \theta_{x}+A_{x,\hat\tau}-i\mu)], \nonumber \end{eqnarray}
and similar expressions for the space hopping and the plaquette interactions. We can then collect 
all the exponentials involving a given $A_{x,\hat\nu}$ and perform the integration over $A_{x,\hat\nu}$. This results in Kronecker deltas relating the various Fourier modes. The final result is 
\begin{eqnarray}
e^{-S_{eff}}&=&\sum_{ \{ m_\Box \} } \left[ \prod_{\Box} I_{m_{\Box}}(\beta_{pl}) \prod_{x}\bigg( I_{n_{x,\hat{s}}}
(2\kappa_s |\phi_x||\phi_{x+\hat{s}}| ) \right. \nonumber \\
&\ &\  \  \  \  \  \times I_{n_{x,\hat{\tau}}}(2\kappa_\tau |\phi_x||\phi_{x+\hat{\tau}}| )\exp(\mu n_{x,\hat{\tau}})\bigg) \bigg], 
\end{eqnarray}
with specific rules to express the link indices $n_{x,\hat\nu}$ in terms of the plaquette indices $m_\Box$ that we now proceed to explain. Given that for $x$ real, $I_n(x)=I_{-n}(x)$ there are several equivalent ways to label the contributions. We use the convention where time is along the vertical axis. 
Starting from the lower left corner of the plaquette and moving counterclockwise, the gauge fields in 
the exponentials come with a plus sign for the first two links and a minus sign for the last two links. For the hopping terms, the gauge fields always come with a positive sign. This results in the rules 
\begin{eqnarray}
\label{eq:nandm1}
n_{x,\hat{s}}&=&m_{below}-m_{above} \\ \label{eq:nandm2}
n_{x,\hat\tau}&=&m_{right}-m_{left} \end{eqnarray} 
where the subscripts such as ``below" refers to the plaquette location with respect to the link. This completely fixes the link indices in terms of the plaquette indices and it is easy to check that Eqs. (\ref{eq:nandm1}-\ref{eq:nandm2}) guarantee that the link indices automatically  satisfy the current conservation imposed by the integration of the $\theta_x$ variables. In other words, the $m_\Box$ are the dual variables \cite{RevModPhys.52.453}. 

Equations (\ref{eq:nandm1}-\ref{eq:nandm2}) have simple electromagnetic analogs. First, $n_{x,\hat\tau}$ can be interpreted as a charge and $m_x$ as an electric field in the spatial direction. 
With this Minkowskian interpretation, Eq.~(\ref{eq:nandm2}) enforces Gauss's law. Second, $n_{x,\hat\nu}$ can be interpreted as a two-dimensional current and $m_x$ as a magnetic field normal 
to the two-dimensional plane. In this Euclidean interpretation, Eqs. (\ref{eq:nandm1}-\ref{eq:nandm2}) express the current as the curl of the magnetic field. A right hand rule can be obtained for the 
following index ordering: time, space, normal direction. A discrete version of Gauss's theorem guarantees that the sum of the charges on a time slice is the last $m$ on the right minus the first $m$ on the left. 

At the lowest order of the strong-coupling expansion we have $\beta_{pl}=0$ and from $I_n(0)=0$ for $n\neq 0$, we see that all the indices must be zeros. 
The effect of the plaquette can be restored perturbatively. This can be organized in an expansion in the 
hopping parameter. In the isotropic case $\kappa_\tau=\kappa_s=\kappa$ we obtain: 
\begin{eqnarray}
&&S_{eff}=\sum_{\langle xy\rangle}\left(-\kappa^2 M_xM_y+\frac{1}{4}\kappa^4(M_xM_y)^2\right) \\
&&-2\kappa^4\frac{I_1(\beta_{pl})}{I_0(\beta_{pl})}\sum_{\Box (xyzw) }M_xM_yM_zM_w +O(\kappa^6)\nonumber
\end{eqnarray}

\section{Monte Carlo calculations}
\label{sec:MC}

Consider the action (\ref{S_u1h}) for the isotropic case $\kappa_\tau=\kappa_s=\kappa$.
We start with the $\beta_{pl}\to\infty$ limit when all gauge variables are frozen
to unity, $U_{x,\hat{\nu}}=1$. In this limit the expectation value of the hopping term
\begin{equation}
L_\phi = \langle{\rm Re}\{\phi_x^\dagger U_{x,\hat{\nu}}\phi_{x+\hat{\nu}}\}\rangle
\end{equation}
can be calculated with the hopping parameter expansion, for small $\kappa$.
It has been derived up to $O(\kappa^5)$ in Ref.~\cite{heitgerphd}. This result
can be generalized to $\beta<\infty$ by including the appropriate factors
of $I_1(\beta)/I_0(\beta)$ for the diagrams that involve plaquettes~\cite{nextpub}.

\begin{figure}
 \includegraphics[width=0.45\textwidth]{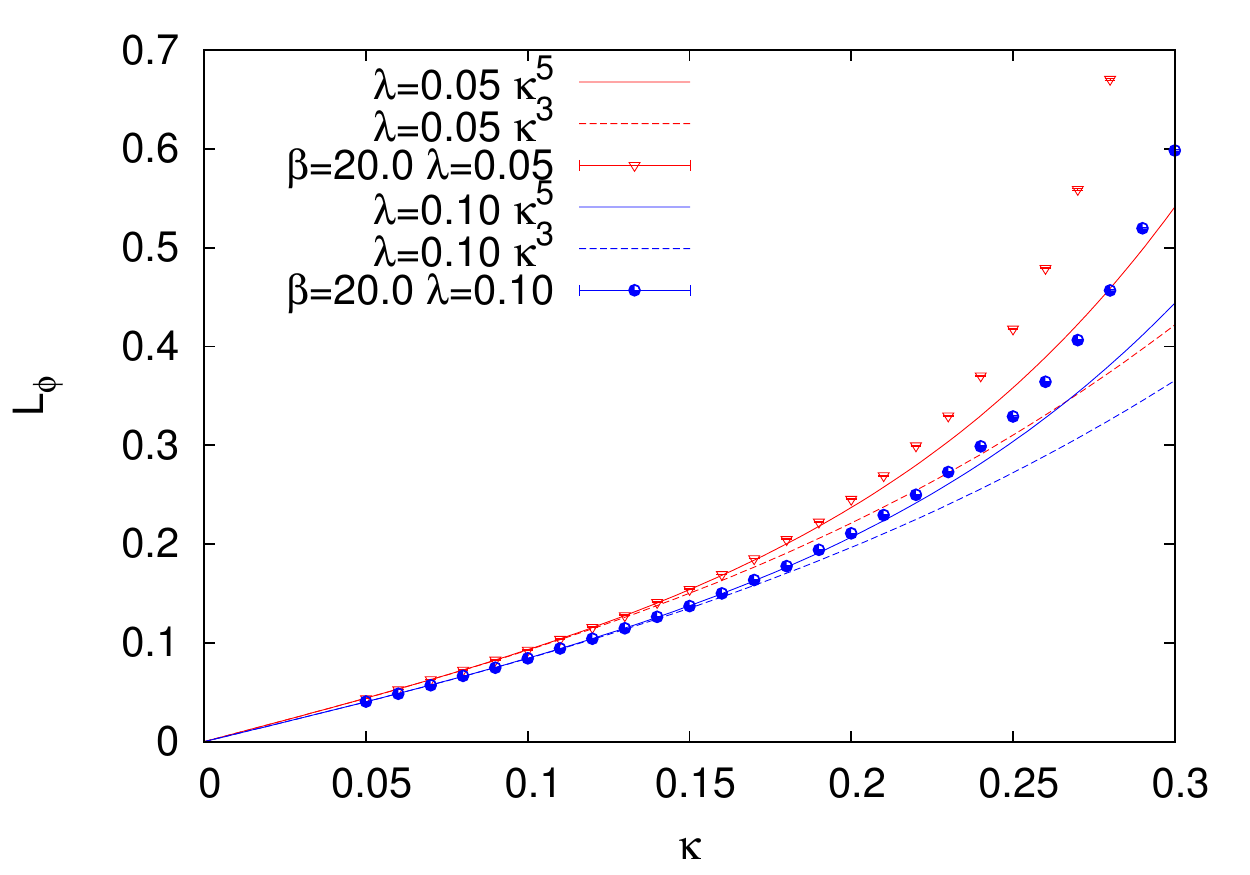}
\caption{$L_\phi$ at $\beta_{pl}=20$ for $\lambda=0.05$ and $\lambda=0.1$ as function
of $\kappa$ compared with the
hopping expansion at $\beta_{pl}=\infty$ at $O(\kappa^3)$ and $O(\kappa^5)$.}
\label{fig_Lf_inf}
\end{figure}
To check the range of validity of the expansion we perform Monte Carlo
simulations at several values of $\beta$, $\kappa$ and $\lambda$ on
a $16^2$ lattice. To test the $\beta\to\infty$ limit we set $\beta=20$ and
for $\lambda=0.05$ and $0.1$ scan the range of $\kappa\in[0.05,0.30]$.
The results for $L_\phi$ are shown in Fig.~\ref{fig_Lf_inf}. The lines represent
the expansion at two orders. The expansion starts to break down around $\kappa=0.15$
at $O(\kappa^3)$ and $\kappa=0.2$ at $O(\kappa^5)$. At the present, exploratory,
stage we use the updating algorithm of Ref.~\cite{Bunk:1994xs} that is applicable
at small self-coupling $\lambda$. Monte Carlo tests of the model
at large $\lambda$ are left for the future.

\begin{figure}
  \includegraphics[width=0.45\textwidth]{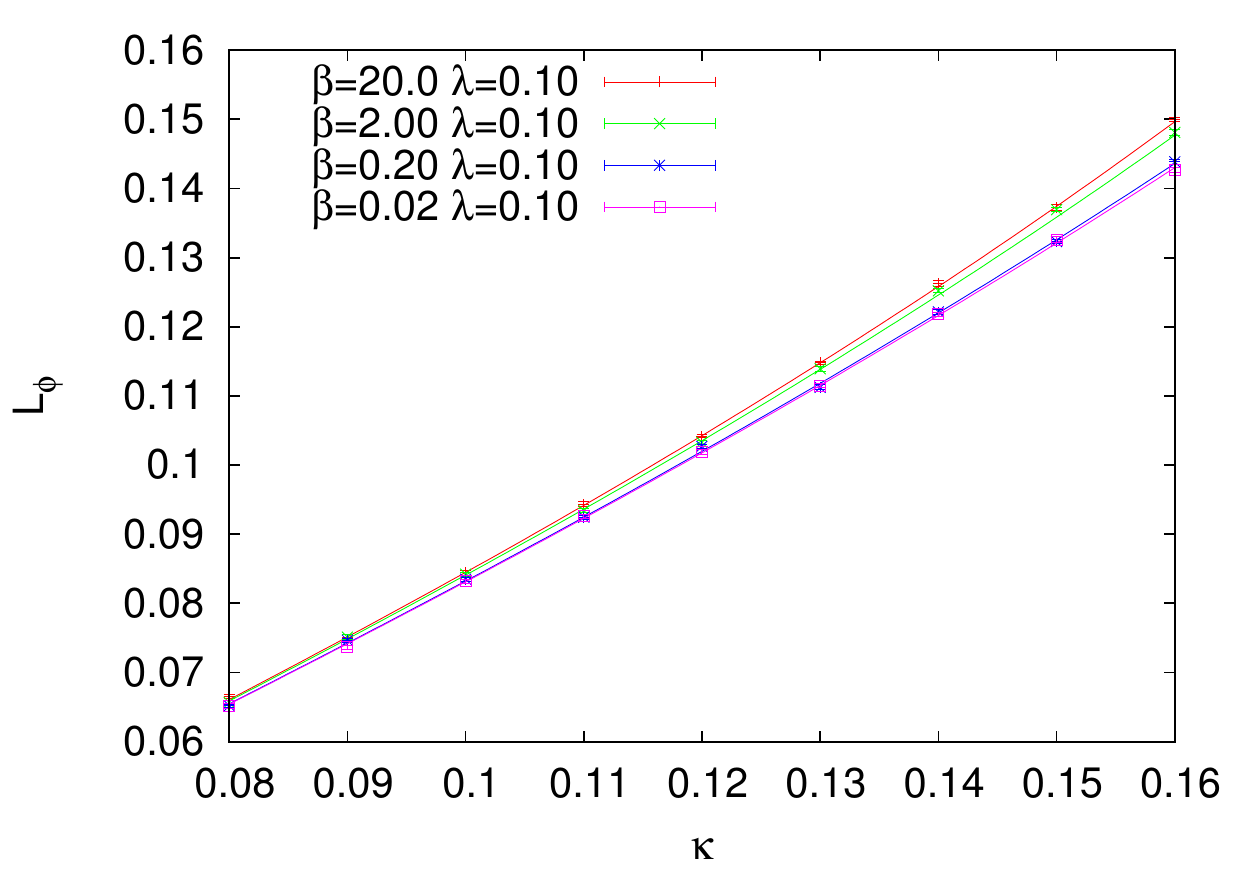}
\caption{$L_\phi$ at $\beta_{pl}=20$, $2$, $0.2$ and $0.02$ for $\lambda=0.1$ as function
of $\kappa$ compared with the
hopping expansion with included dependence on $\beta_{pl}$ up to $O(\kappa^5)$.}
\label{fig_Lf_beta}
\end{figure}
To study the dependence on $\beta$ we focus on the $\kappa\in[0.08,0.16]$ range at
$\lambda=0.1$ and perform calculations at $\beta_{pl}=20$, $2$, $0.2$ and $0.02$.
The results for $L_\phi$ are shown in Fig.~\ref{fig_Lf_beta}.

\begin{figure}
  \includegraphics[width=0.45\textwidth]{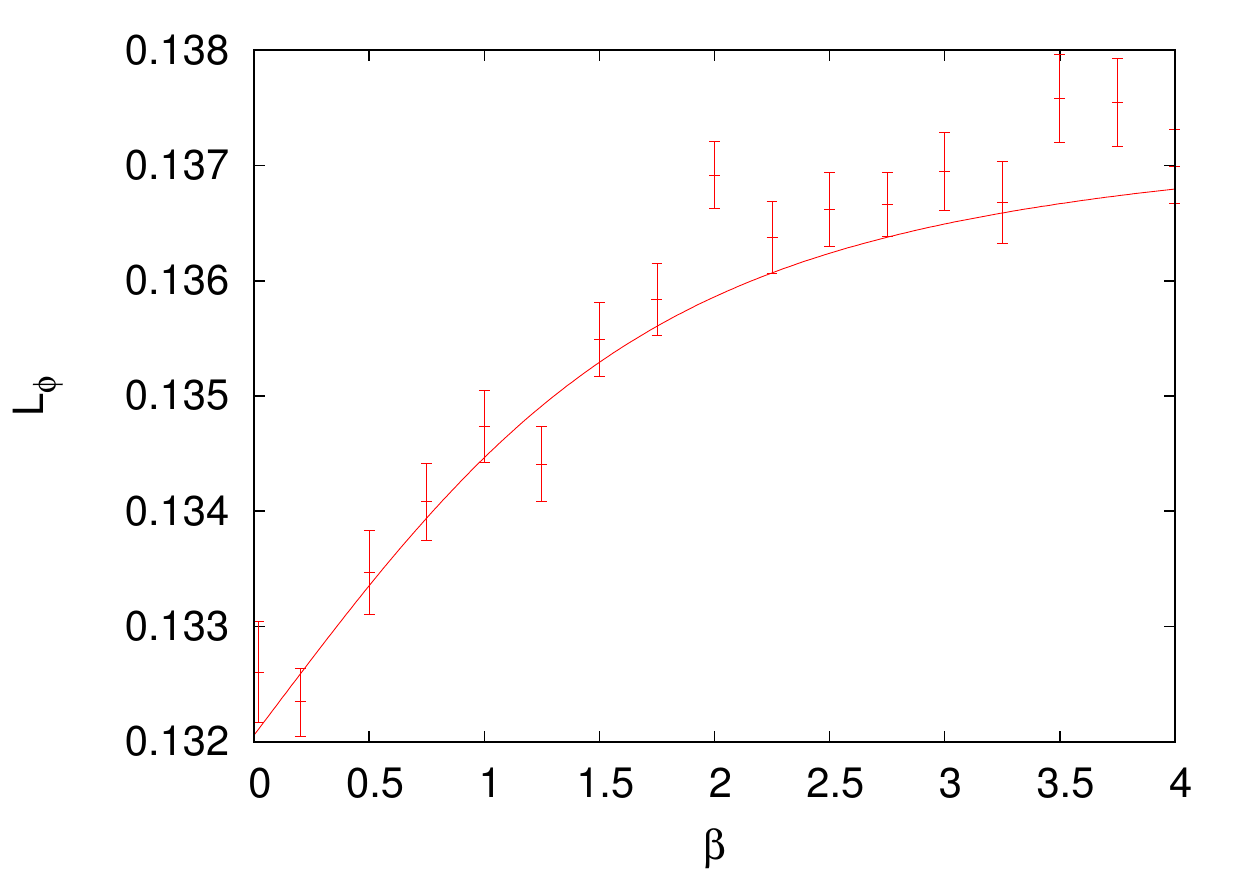}
\caption{$L_\phi$ at fixed $\kappa=0.15$ as function of $\beta_{pl}$ for $\lambda=0.1$ compared with the
hopping expansion with included dependence on $\beta_{pl}$ up to $O(\kappa^5)$.}
\label{fig_Lf_kappa}
\end{figure}
To understand the dependence on $\beta_{pl}$ better we also calculate $L_\phi$ for several
values of $\beta_{pl}$ at fixed $\kappa=0.15$. The results together with the
hopping expansion are shown in Fig.~\ref{fig_Lf_kappa}.

\section{The large $\lambda$ limit}
\label{sec:lambda}
 We now turn to the limit where $\lambda$ becomes arbitrarily large. In this limit, $M_x$ is frozen to 1, or in other words, the 
 Brout-Englert-Higgs mode becomes infinitely massive. We are then left with compact variables of integration in the original formulation ($\theta_x$ and $A_{x,\hat\nu}$) and the Fourier expansions described before lead to expressions of the partition function in terms of discrete sums. 
 
 As explained in Ref. \cite{PhysRevD.88.056005}, these sums can be formulated in a compact way using tensorial notations. 
 In order to simplify the equations below we define
\begin{equation}
    t_{n}(z) = \frac{I_{n}(z)}{I_{0}(z)} .
\end{equation}
These normalized Bessel functions have useful properties. For instance $t_n(0)=\delta_{n,0}$. 
For $z$ non-zero and finite, we have 
\beq1=t_0(z)>t_1(z)>t_2(z)>\dots>0.\enq In addition, for sufficiently large $z$,
\beq
\label{eq:asbess}
t_n(z)\approx 1-\frac{n^2}{2z}.
\enq
Following the general principles of the construction \cite{PhysRevD.88.056005}, we attach a $B^{(\Box)}$ tensor  to every plaquette 
 \begin{eqnarray}
            \label{eq:ttensorB}
             &&B^{(\Box)}_{m_1m_2m_3m_4}\\
            =&&\begin{cases} 
                t_{m_\Box}(\beta_{pl}),  &\mbox{if } m_1=m_2=m_3=m_4=m_\Box \\ \nonumber
                0, & \mbox{otherwise},
            \end{cases}
        \end{eqnarray}
a $A^{(s)}$ tensor to the horizontal links
\beq
            A^{(s)} _{m_{above}m_{below}}=t_{|m_{below}-m_{above}|}(2\kappa_s), 
            \enq     
 and a $ A^{(\tau)}$ tensor to the vertical links        
             \beq
            A^{(\tau)} _{m_{left}m_{right}}=t_{|m_{left}-m_{right}|}(2\kappa_\tau)\ {\rm e}^{(m_{right}-m_{left})\mu}.
            \enq              

\begin{figure}
    \includegraphics[width=0.4\textwidth]{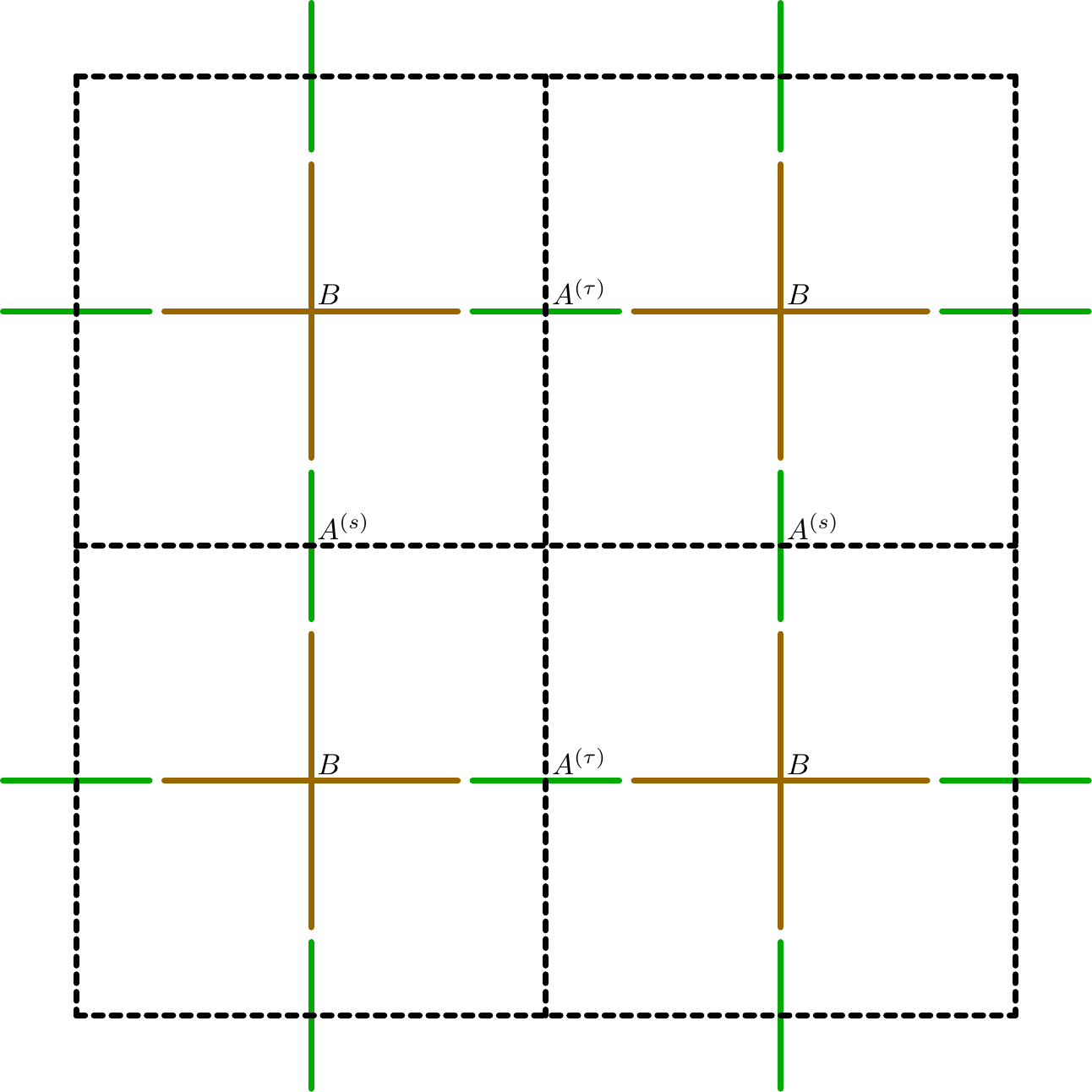}
       \caption{The basic $B$ and $A$ tensors (in brown and
    green, respectively, colors online).  The $A^{(s)}$ are associated with the vertical
    tensors, and the horizontal (spatial) links of the lattice.  The $A^{(\tau)}$
    are associated with the horizontal tensors, and the vertical (temporal)
    links of the lattice.}
     \label{fig:basictensors}
\end{figure}

 The partition function can now be written as
        \begin{eqnarray}
         &&   Z= ( I_{0}(\beta_{pl})I_0(2\kappa_s)I_0(2\kappa_\tau))^V \times \\
          &&  {\rm Tr}\left[ \prod_{h,v,\Box}A^{(s)} _{m_{above}m_{below}} A^{(\tau)} _{m_{left}m_{right}}B^{(\Box)}_{m_1m_2m_3m_4}\right].\nonumber
        \end{eqnarray}
The traces are performed by contracting the vertical and horizontal indices as shown in Fig.~\ref{fig:basictensors}.
Note that the tensor $A^{(s)}$ associated with a horizontal space link is represented by a vertical line orthogonal to the link and  
the tensor $A^{(\tau)}$ associated with a vertical time link is represented by a horizontal line orthogonal to the link.

The traces can also be expressed in terms of a transfer matrix $\mathbb{T}$ which can be constructed in the following way.
First, we define a matrix $\mathbb{B}$ as the product of the plaquette tensors on a time slice alternating with the link tensor corresponding to the vertical links 
in between the plaquettes. There are two natural ways to impose boundary conditions. The first is to connect the last $A$ tensor with the first $B$ tensor (periodic boundary conditions), the second is to impose $m=0$ for the first and last $B$ tensor (open boundary conditions). However, in both cases, the total charge on the interval 
has to be zero. A more general option is to allow arbitrary $m$'s at each end. This would allow us to consider appropriately selected charge sectors. In the 
following we will focus on the open boundary conditions and define 
 \begin{eqnarray}
&&\mathbb{B}_{(m_1,m_2,\dots m_{N_s})(m_1',m_2'\dots m_{N_s}')}=t_{m_1}(2\kappa_\tau)\delta_{m_1,m_1'}t_{m_1}(\beta_{pl})\times\nonumber\\
&& t_{|m_1-m_2|}(2\kappa_\tau)\delta_{m_2,m_2'}t_{m_2}(\beta_{pl})t_{|m_2-m_3|}(2\kappa_\tau)\dots \nonumber \\
&&t_{m_{N_s}}(\beta_{pl})t_{m_{N_s}}(2\kappa_\tau).
\end{eqnarray}
Note that with this choice of open boundary conditions, the chemical potential has completely disappeared. If we had chosen different $m$'s at the end 
allowing a total charge $Q$ inside the interval, we would have an additional factor $\exp (\mu Q)$. 
We next define a matrix $\mathbb{A}$ as the product
\begin{eqnarray}
&&\mathbb{A}_{(m_1,m_2,\dots m_{N_s})(m_1',m_2'\dots m_{N_s}')}=\\
&&t_{|m_1-m_1'|}(2\kappa_s)t_{|m_2-m_2'|}(2\kappa_s)\dots t_{|m_{N_s}-m_{N_s}'|}(2\kappa_s).\nonumber
\end{eqnarray}

With these notations we can construct a symmetric transfer matrix $\mathbb{T}$. Since $\mathbb{B}$ is diagonal, real and positive, we can define its square root in an obvious way and write the transfer matrix as 
\beq
\label{eq:tm}
\mathbb{T}=\sqrt{\mathbb{B}}\mathbb{A}\sqrt{\mathbb{B}}.
\enq
With this definition, the partition function can be written as
\beq Z= ( I_{0}(\beta_{pl})I_0(2\kappa_s)I_0(2\kappa_\tau))^V {\rm Tr}\left[\mathbb{T}^{N_\tau}\right].
\enq
Alternatively, we could diagonalize the symmetric matrix $\mathbb{A}$ and define the (dual) transfer matrix 
\beq
\tilde{\mathbb{T}}=\sqrt{\mathbb{A}}\mathbb{B}\sqrt{\mathbb{A}}.
\enq

\begin{figure}
    \includegraphics[width=0.4\textwidth]{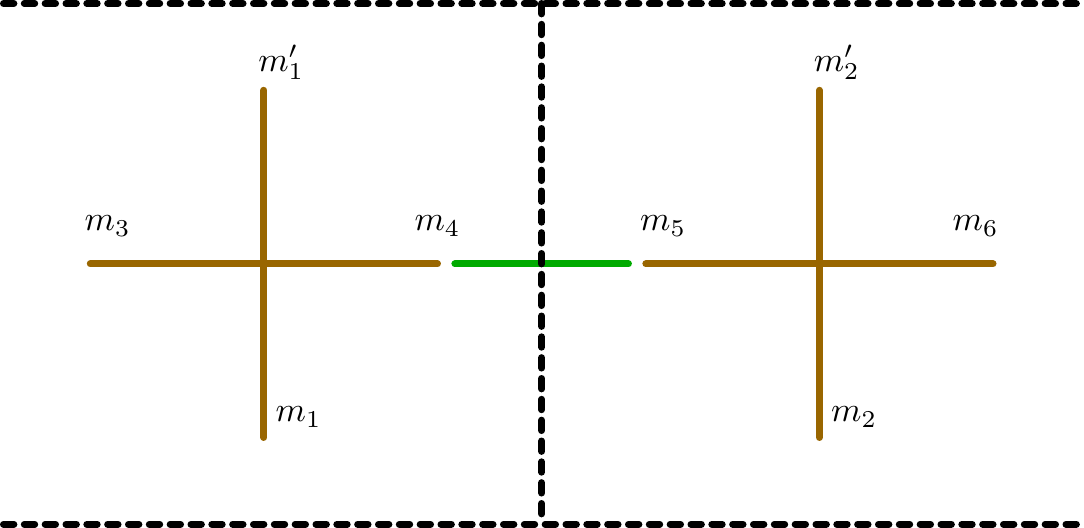}
     \caption{Part of the construction of the blocked $B'$ tensor.  This shows
    the contraction of the $B$ and $A^{(\tau)}$ tensors.  The dashed
    lines are the links of the original lattice.}
\label{fig:topandbot}
\end{figure}

The $\mathbb{A}$ and $\mathbb{B}$ matrices can be constructed by a recursive blocking methods similar 
to those discussed in Ref. \cite{PhysRevD.88.056005}.  
We can construct a new $B$ tensor, called $B'$, by contracting two $B$
tensors on both sides of a $A^{(\tau)}$ tensor.  This process can be iterated and is illustrated 
in Fig.~\ref{fig:topandbot}.  Using tensorial notation we write
\begin{eqnarray}
  &&  B'_{m_3m_6 M(m_1, m_2) M'(m_1',m_2')} =\\\
 &&   \sum_{m_4 , m_5} B_{m_3 m_4 m_1 m_1'} A^{(\tau)}_{m_4 m_5} B_{m_5 m_6 m_2 m'_2 },\nonumber
\end{eqnarray}
where the notation $M(m_1, m_2)$ stands for the product state, $M = m_1 \otimes m_2$.
One can continue blocking horizontally as above until the desired spatial size
is achieved resulting in the matrix $\mathbb{B}$.  By their very nature
the $B$'s enforce that this object is diagonal in the collective product-state
 indices of
the upper and lower tensor legs.  
For practical reasons, truncation methods need to be introduced at the beginning and after each blocking 
\cite{PhysRevB.86.045139,PhysRevB.87.064422}.

\begin{figure}
       \includegraphics[width=0.3\textwidth]{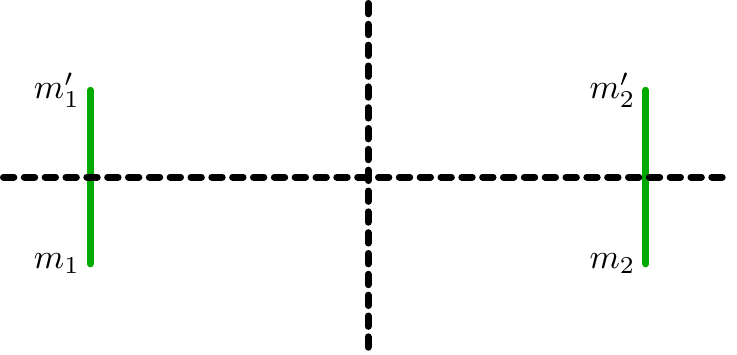}
        \caption{Graphical representation of the blocking of the $A$ tensors. The vertical
    tensors are the $A^{(s)}$ and the dashed
    lines are the links of the original lattice.}
    \label{fig:amatrix}
\end{figure}

Now consider the $A^{(s)}$ matrices which correspond to the
horizontal links of the original lattice as in  Fig. \ref{fig:amatrix}.  If one takes their outer product and 
collects the upper indices into a single product-state index, and does the same for
the lower indices, one has another matrix built out of the vertical $A^{(s)}$ 
tensors.  Once again in tensorial notation this can be written as
\begin{equation}
    A^{'(s)}_{M(m_1, m_2) M'(m_1',m_2')} = A^{(s)}_{m_1 m_1'} A^{(s)}_{m_2 m'_2}.
\end{equation}
One can continue taking the product of $A^{(s)}$ matrices until the desired
spatial size has been reached resulting in the matrix $\mathbb{A}$.

\section{The limit $\beta_{pl} \rightarrow\infty$}
\label{sec:weak}

In this section, we discuss the case where both $\lambda$ and $\beta_{pl}$ are infinite. In this limit, the correspondence between the classical lattice model and a two-species 
Hubbard model has been outlined in Ref. \cite{PhysRevA.90.063603} and we will show that the method discussed in Sec. \ref{sec:lambda} allows us to match the spectra at small volume with very good accuracy. This section is a warm-up for the more complicated situations discussed in Secs. \ref{sec:time} and \ref{sec:bh2}. 

In the limit where $\beta_{pl}$ becomes arbitrarily large, we need to maximize ${\rm Re}U_{pl,x}$. This is accomplished when $A_{x,\hat\nu}$ is a gauge transform of 0. In other words, we can set $A_{x,\hat\nu}=0$ and 
the gauge equivalent configurations are taken into account by integrating over the Nambu-Goldstone modes $\theta_x$. We recover the familiar $O(2)$ model. As explained in Ref. \cite{PhysRevD.17.2637,RevModPhys.51.659,Polyakov:1987ez}, the time continuum limit can be achieved by taking the limit 
$\kappa_\tau\rightarrow \infty$ while keeping the product $\kappa_\tau\kappa_s$ constant. This leads to a  Hamiltonian for quantum rotors located at each spatial site and having quantized angular momentum running over positive and negative integers. In order to realize this Hamiltonian on optical lattices, we considered  \cite{PhysRevA.90.063603}  a spin-1 truncation where the angular momentum at each site is restricted to the values 0 and $\pm1$. This truncation has very small effects on the phase diagram provided that the hopping parameter and the chemical potential are not too large. 

Following Refs. \cite{PhysRevA.90.063603,Bazavov:2014lla}  with the replacements $\beta_\nu\rightarrow 2\kappa_\nu$, the Hamiltonian for the spin-1 approximation of the $O(2)$ model reads:
\begin{eqnarray}
&&\hat{H}=\frac{\tilde{U}}{2}\sum_i \left(\hat{L}^z_{(i)}\right)^2
\label{eq:rotor2}\\
&&-\tilde{\mu}\sum_i \hat{L}^z_{(i)}
-\frac{\tilde{J}}{4}\sum_{i}
\left(\hat L^+_{(i)}\hat L^-_{(i+1)}+\hat L^-_{(i)}\hat L^+_{(i+1)}\right)\nonumber \ ,
\end{eqnarray}
with the now dimensionful quantities 
\beq
\tilde{U}\equiv\frac{1}{2\kappa_\tau a}, \ \tilde{\mu}\equiv\frac{\mu}{a},\  \tilde{J}\equiv\frac{2\kappa_s}{a},\enq and $a$  the time lattice spacing. 
If we use open boundary conditions, the sums run over all spatial sites in the first two sums 
and all but the last in the third sum. 
This approximate Hamiltonian can be matched \cite{PhysRevA.90.063603,Bazavov:2014lla}  with the effective Hamiltonian of  a Bose-Hubbard model discussed in sec. \ref{sec:bh2}, that, we expect, can be implemented on optical lattices. 

\begin{figure}[h]
 \includegraphics[width=2.4in]{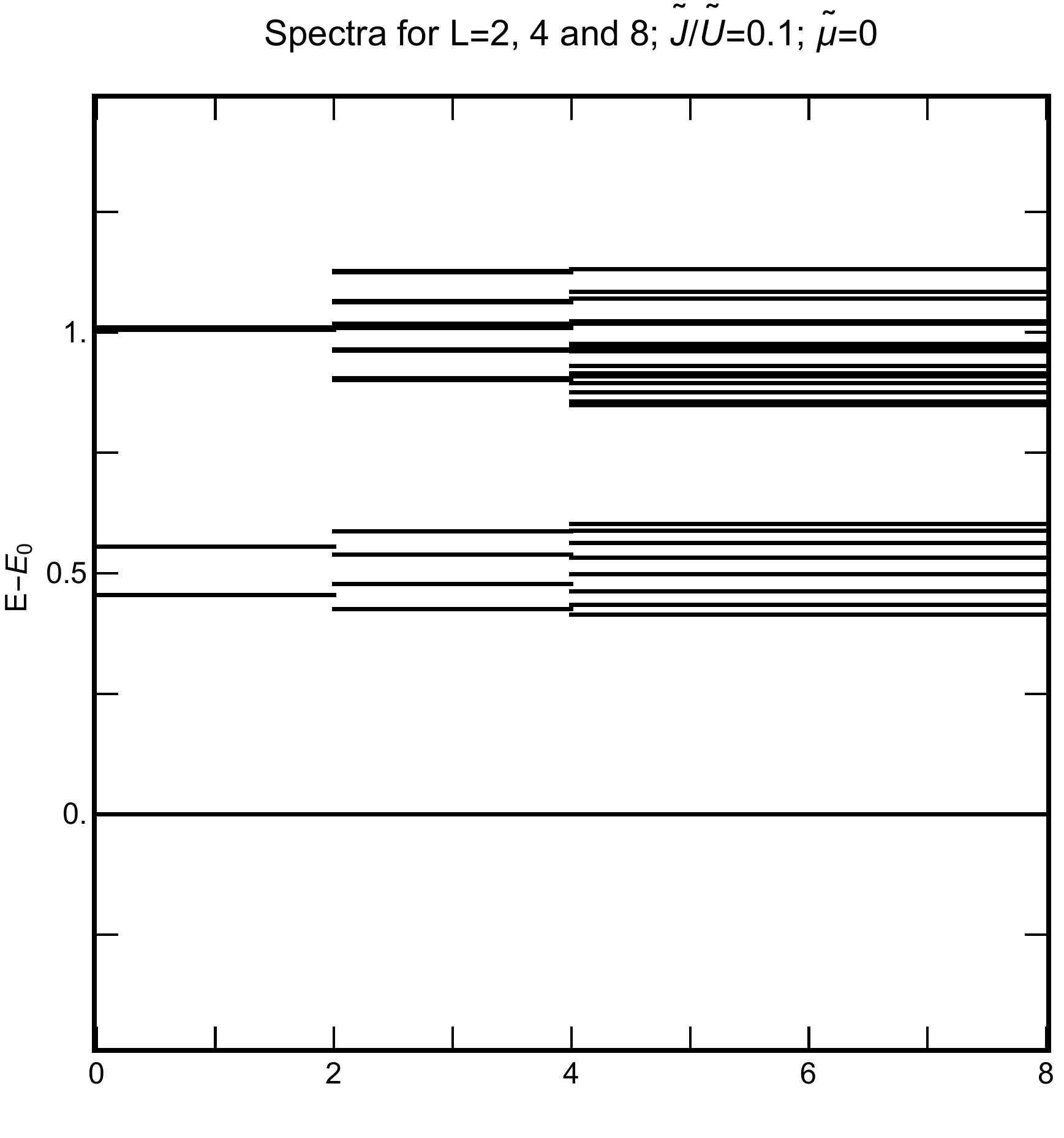}
\caption{ \label{fig:o2s}$O(2)$ spectra in $\tilde{U}$ units for $L$=2, 4, and 8, with $\tilde{J}$=0.1, $\tilde{\mu}=0$. Some higher energy states are not shown on the figure.}
\end{figure}

The first step is to understand the matching in the simple situation where $\tilde{U}$ is large compared to $\tilde{\mu}$ and $\tilde{J}$.
In this case, $\tilde{U}$ sets the scale of the mass gap and $\tilde{\mu}$ and $\tilde{J}$ introduce small splittings. We now use energy units where $\tilde{U}$=1. 
The time continuum limit can be obtained numerically as we keep blocking the transfer matrix in the space direction as discussed in Sec. \ref{sec:lambda}. 
The spectra in blocks of size 2, 4 and 8 in the spin-1 approximation are shown in Fig. \ref{fig:o2s} for $\tilde{J}$=0.1, $\tilde{\mu}=0$. A nonzero chemical potential would 
split the charge conjugated states and make the graph difficult to read. An example will be shown for $L=2$ in Sec. \ref{sec:bh2}.

\section{The time continuum limit and the energy spectrum}
\label{sec:time}

In this section, we construct the time continuum limit of the transfer matrix $\mathbb{T}$ defined by Eq. (\ref{eq:tm}) starting now with the more general situation where $\beta_{pl}$ is finite. 
In order to obtain 
Hamiltonians corresponding to Bose-Hubbard models, 
a guiding strategy is to assume that the spatial hopping $\kappa_s$ is proportional to some hopping energy. In the limit $\kappa_s=0$, the inside matrix $\mathbb{A}$  becomes the identity and the transfer matrix $\mathbb{T}=\sqrt{\mathbb{B}}\mathbb{A}\sqrt{\mathbb{B}}$ becomes $\mathbb{B}$ which is diagonal. In this limit, the only way to obtain a time continuum limit, in other words to have $\mathbb{T}$ close to the identity, is to require that {\it both} $\kappa_\tau$ and $\beta_{pl}$ become  large. At leading order in the inverse of these large parameters,  the eigenvalues of $\mathbb{T}$ are 
\begin{eqnarray}
\label{eq:eig}
&&\lambda_{(m_1,m_2,\dots m_{N_s})}= \\ &&1-\frac{1}{2}[(\frac{1}{\beta_{pl}} (m_1^2+m_2^2+\dots+m_{N_s}^2)+\nonumber\\
&&\hskip40pt \frac{1}{2\kappa_\tau}(m_1^2+(m_2-m_1)^2+\dots \nonumber \\
 && \hskip80pt\dots+(m_{N_s}-m_{N_s-1})^2+m_{N_s}^2)]\nonumber
\end{eqnarray}
There are two limiting situations: $1<<\beta_{pl}<<\kappa_\tau$ and $1<<\kappa_\tau<<\beta_{pl}$. 
In the first case, $1/\beta_{pl}$ is the largest coefficient and the low energy states are those with a few $m_i$ nonzero and consequently only a few differences 
of $m$'s are nonzero and the second term in Eq. (\ref{eq:eig}) is a perturbation. In the second case, $1/\kappa_\tau$ is the largest coefficient and we could 
guess that to minimize its contribution, we need to take most of the $m$'s to be equal and just create a few charges. For a large volume, this can cause the first term of Eq. (\ref{eq:eig}) to dominate, which signals an infrared instability. In electrostatic terms, creating one charge also creates a constant electric field over the entire volume. 
The linear potential is responsible for the confinement and trying to treat the interaction perturbatively results in infrared problems. 

For this reason, in the following, we will 
only consider the case $1<<\beta_{pl}<<\kappa_\tau$ and set the scale with the (large) gap energy
\beq
\tilde{U}_P\equiv \frac{1}{a\beta_{pl}}\ .
\enq
It is important to distinguish this scale from the $\tilde{U}$ introduced in Sec. \ref{sec:weak}.
$\tilde{U}_P$ is associated with the plaquette quantum number $m$ (the energy to create a flux tube across one lattice spacing between the opposite charges), while $\tilde{U}$ in Sec. \ref{sec:weak} is associated with the quantum number $n$  (the energy to create a single charge). 
In addition, we define the (small) energy scales
\beq
\tilde{Y}\equiv \frac{1}{2\kappa_\tau a} =\frac{\beta_{pl}}{2\kappa_\tau} \tilde{U}_P, \  \enq
which plays the same role as $\tilde{U}$ in Sec. \ref{sec:weak} and the space hopping parameter
\beq
\tilde{X}\equiv\sqrt{2}\beta_{pl}\kappa_s\tilde{U}_P.
\enq

We are now in position to derive an expression for the Hamiltonian in the spin-1 approximation where the plaquette quantum number $m$ takes values $\pm1$ and 0 only. 
The effect of $\kappa_s$ can be studied by linearization. In the case of two spins, we find that 
\beq
\partial \mathbb{T}/\partial \kappa_s |_{\kappa_s=0}=\sqrt{2}(\bar{L}_{(1)}^x+\bar{L}_{(2)}^x)
\enq
We use the notation $\bar{L}_{(1)}^x$ to denote the first generator of the spin-1 rotation algebra at the site $(1)$. The notation $\bar{L}$ is used to emphasize that the spin 
is related to the $m$ quantum numbers attached to the plaquettes in contrast to the spin-1 generators $\hat{L}$ in Sec. \ref{sec:weak} having a spin related to the charges $n$ attached to the time links. The final form of the Hamiltonian $\bar{H}$ for $1<<\beta_{pl}<<\kappa_\tau$ is 
\begin{eqnarray}
&&\bar{H}=\frac{\tilde{U}_P}{2}\sum_i \left(\bar{L}^z_{(i)}\right)^2+
\label{eq:ham} \\
&&\frac{\tilde{Y}}{2} {\sum_i}  ' (\bar{L}^z_{(i)}-\bar{L}^z_{(i+1)})^2-
\tilde{X}\sum_{i}
\bar L^x_{(i)}\nonumber \ ,
\end{eqnarray}
where $\sum_i '$ is a short notation to include the single terms at the two ends as in Eq. (\ref{eq:eig}), \textit{i.e.} besides $(\bar{L}^z_{(1)}-\bar{L}^z_{(2)})^2$, $(\bar{L}^z_{(2)}-\bar{L}^z_{(3)})^2$, ... , $(\bar{L}^z_{(N_s-1)}-\bar{L}^z_{(N_s)})^2$ terms this sum contains $(\bar{L}^z_{(1)})^2$ and $(\bar{L}^z_{(N_s)})^2$.

The process outlined in this paper can in principle be generalized for non-Abelian gauge groups in 1+1 and higher
spatial dimensions. Tensor formulations and exact blocking procedures have already been worked out
for non-Abelian gauge groups in Ref.~\cite{PhysRevD.88.056005}. These questions are under active consideration.

\section{A two-species Bose-Hubbard model implementation}
 \label{sec:bh2}

In Refs. \cite{PhysRevA.90.063603,Bazavov:2014lla}, we have proposed to match the Hamiltonian of the $O(2)$ model given in Eq. (\ref{eq:rotor2}) with the  two-species Bose-Hubbard Hamiltonian on a square optical lattice 
\begin{eqnarray}
&\mathcal{H}&=-\sum_{\langle ij\rangle}(t_a a^\dagger_i a_j+t_b b^\dagger_i b_j+h.c.)-\sum_{i, \alpha}(\mu_{a+b}+\Delta_\alpha)n^\alpha_i\nonumber\\
&+&\sum_{i, \alpha}\frac{U_\alpha}{2}n^\alpha_i(n^\alpha_i-1)+W\sum_in^a_in^b_i+\sum_{\langle ij\rangle\alpha}V_\alpha n^\alpha_in^\alpha_j
\end{eqnarray}
with  $\alpha=a,b$ indicating two different species and with $n^a_i=a^\dagger_ia_i$ and $n^b_i=b^\dagger_ib_i$. 
In this expression, the chemical potential $\mu_{a+b}$ is associated with the conservation of $n^a+n^b$ and should not be confused with the chemical potential 
introduced in the previous section which couples to $n^a-n^b$ and breaks the charge conjugation symmetry. 
In the limit where $U_a=U_b=U$ and $W$ and $\mu_{a+b}=(3/2) U$ are much larger than any other energy scale, 
we have the condition $n^a_i+n^b_i=2$
for the low energy sector. 
The three states 
 $|2,0\rangle$, $|1,1\rangle$ and $|0,2\rangle$ satisfy this condition 
 and correspond to the three states of the spin-1 projection considered above.

Using degenerate perturbation theory \cite{Kuklov:2003bra,Kuklov:2004hl}, we found \cite{PhysRevA.90.063603,Bazavov:2014lla} that 
\begin{eqnarray}\label{eq:H_eff}
\mathcal{H}_{eff}&=&
\left(\frac{V_a}{2}-\frac{t_a^2}{U_0}+\frac{V_b}{2}-\frac{t_b^2}{U_0}\right)\sum_{\langle ij\rangle}L^z_iL^z_j\nonumber \\&+&
\frac{-t_at_b}{U_0}\sum_{\langle ij\rangle}(L^+_iL^-_j+L^-_iL^+_j)+(U_0-W)\sum_{i}(L^z_i)^2\nonumber\\
&+&\left[\left(\frac{pn}{2}V_a+\Delta_a-\frac{p(n+1)t_a^2}{U_0}\right)-\left(\frac{pn}{2}V_b\right.\right.\nonumber\\
&+&\left.\left.\Delta_b-\frac{p(n+1)t_b^2}{U_0}\right)\right]\sum_{i}L^z_i,
\end{eqnarray}
where $p$ is the number of neighbors and $n$ is the occupation ($p=2$, $n=2$ in the case under consideration).
$\hat L$ is the angular momentum operator in the representation $n/2$.
The effective hamiltonian (\ref{eq:H_eff}) without the last term has been
studied in~\cite{PhysRevB.65.220511} and shows a rich phase diagram.

\begin{figure}[h]
 \includegraphics[width=2.in]{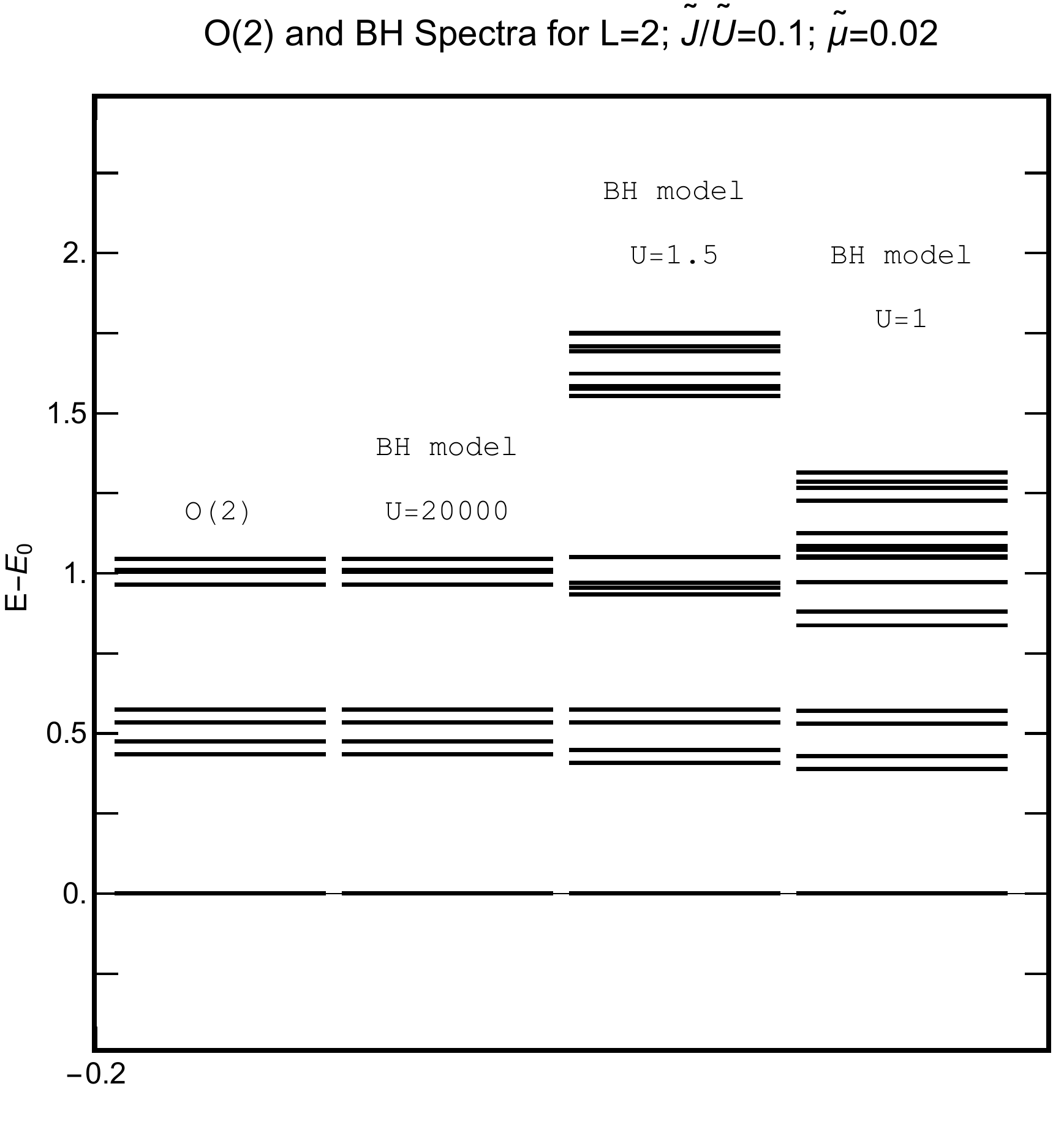}
 \includegraphics[width=2.in]{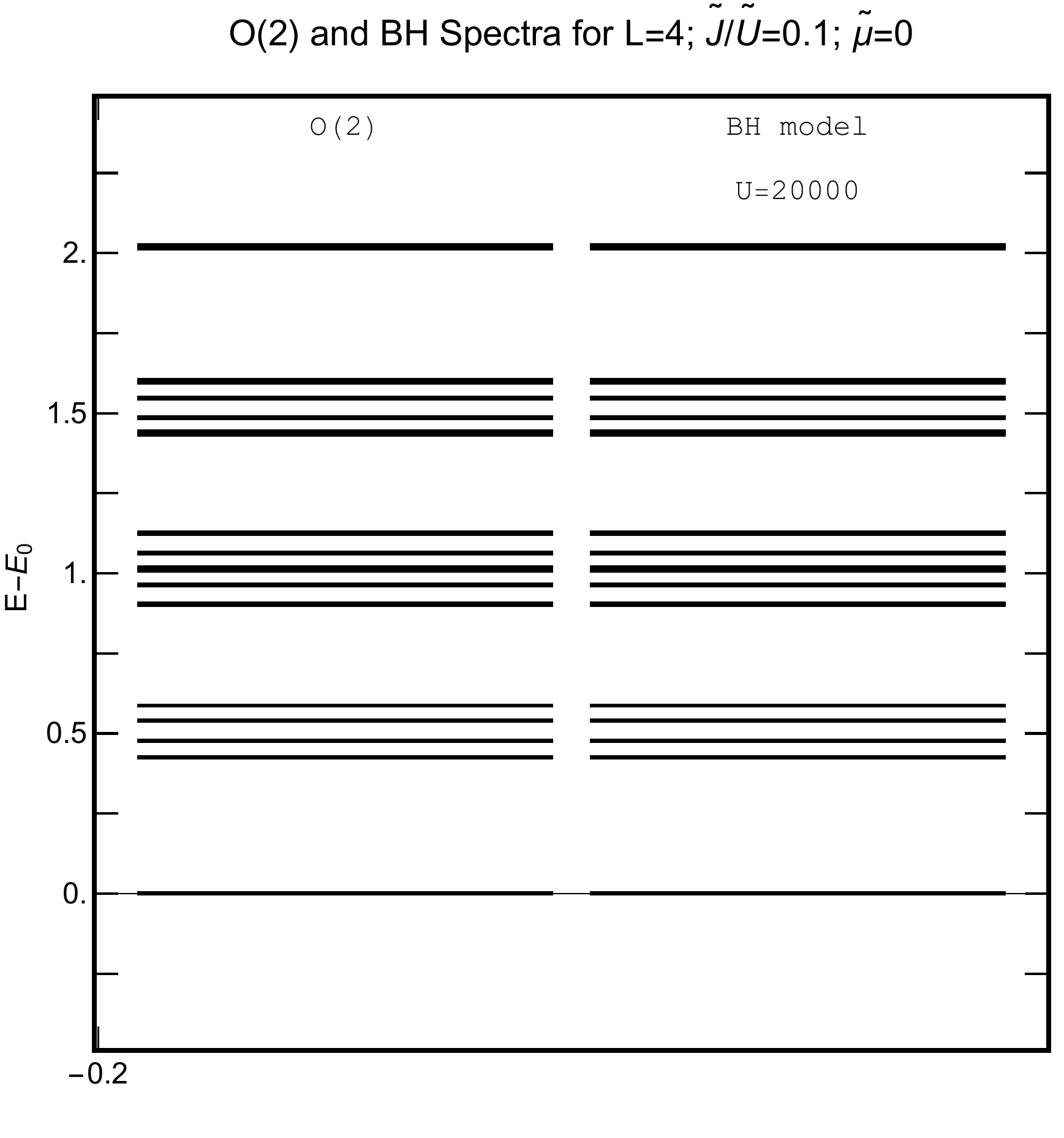}
\caption{\label{fig:mo2} $O(2)$ with $\tilde{J}/\tilde{U} =0.1$ and Bose-Hubbard spectra for $L=2$ with $\tilde{\mu}=0.02$ (top) and $L=4$ with $\tilde{\mu}=0$ (bottom). }
\end{figure}

In order to match with the $O(2)$ model, we need to tune 
the hopping amplitude as  $t_\alpha=\sqrt{V_\alpha U/2}$ and have 
$\tilde{J}=4\sqrt{V_aV_b}$, $\tilde{U}=2(U-W)$, and $\tilde{\mu}=-(\Delta_a-V_a)+(\Delta_b-V_b)$. 
The matching is illustrated in Fig. \ref{fig:mo2} for $\tilde{J}/\tilde{U} =0.1$ and large $U$. In the limit $U\rightarrow \infty$, the matching is excellent. 
As we lower $U$, high energy states get a lower energy and can mix with the low energy states. These states may be related to 
the modes that become infinitely massive when $\lambda\rightarrow\infty$. 

The optical lattice implementation is discussed in Ref. \cite{PhysRevA.90.063603} for
a $^{87}$Rb and $^{41}$K Bose-Bose mixture where an interspecies Feshbach resonance is accessible. 
The interspecies interaction ($W$) can therefore be controlled by an external magnetic field. 
 The extended interaction, $V_\alpha$, is present and small when we consider Wannier gaussian wave functions sitting on nearby lattice sites~\cite{2006PhRvA..73a3625M}. 
 
\begin{figure}[h]
 \includegraphics[width=2.in]{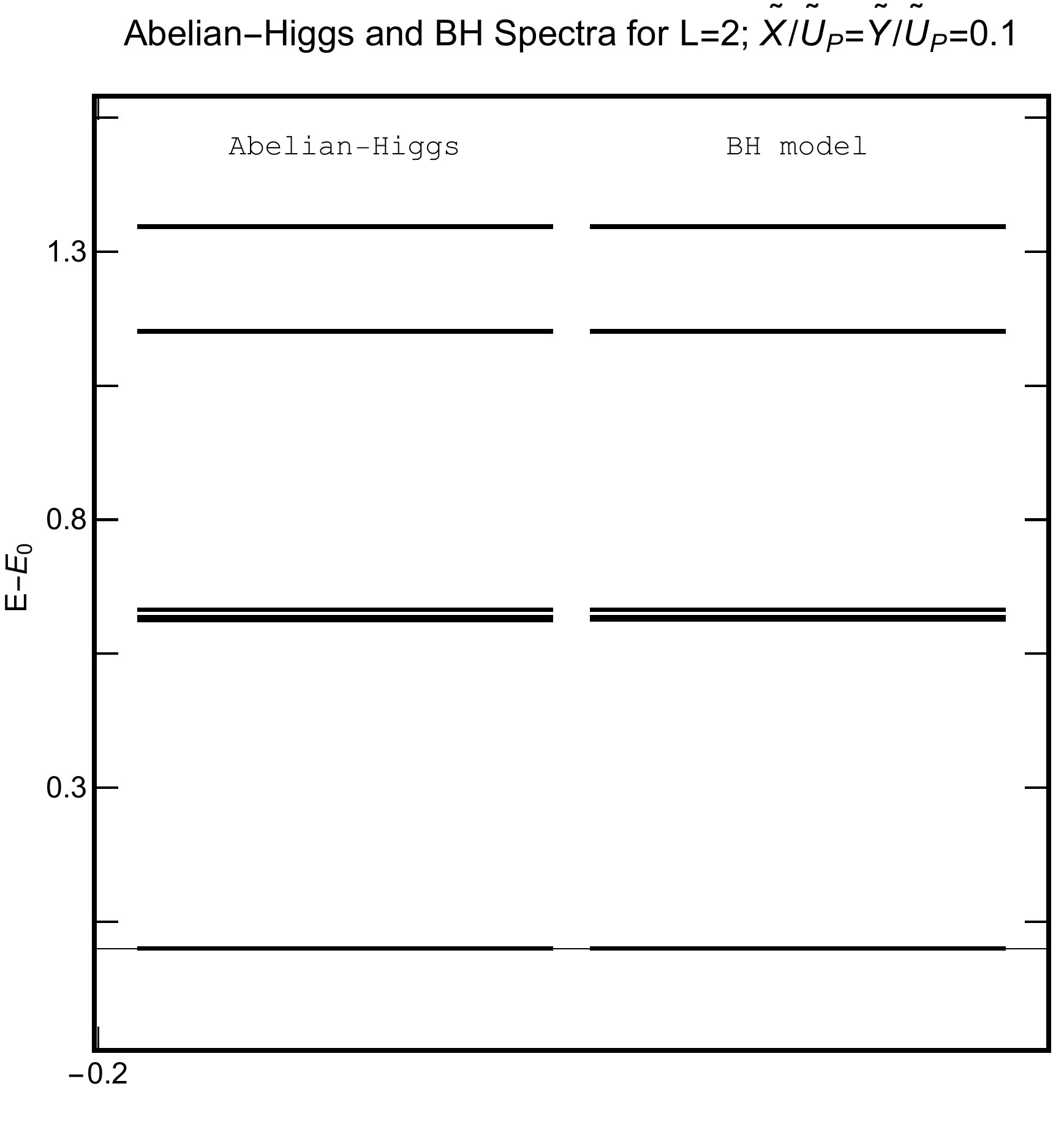}
  \includegraphics[width=2.in]{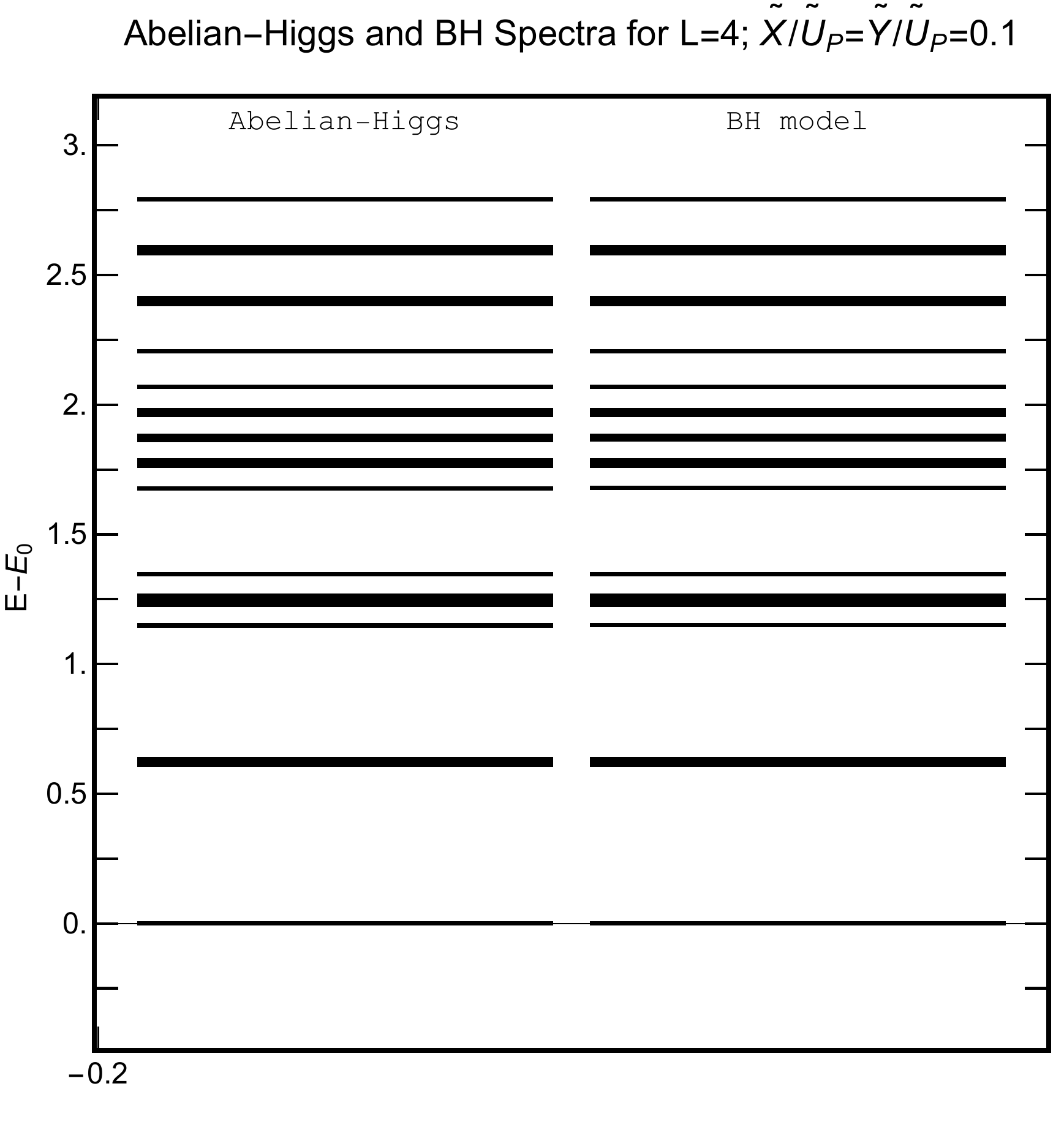}
\caption{\label{fig:hbh4} Abelian-Higgs model with $\tilde{X}/\tilde{U} =0.1$, $\tilde{Y}/\tilde{U} =0.1$ and the corresponding Bose-Hubbard spectra for $L=2$ (top) and $L=4$ (bottom).}
\end{figure}

 For the Hamiltonian $\bar{H}$ in Eq. (\ref{eq:ham}) corresponding to $1<<\beta_{pl}<<\kappa_\tau$, we need to introduce a new interaction to represent the 
 $\bar{L}^x$ effects that interchange the $m=0$ states with the $m=\pm1$ states. 
 This can be achieved by adding the piece 
 \beq
 \Delta H=-\frac{t_{ab}}{2}\sum_i  (a^\dagger_i b_i+  b^\dagger_i a_i)\ .
 \enq
 The matching between the two models can be achieved by imposing $t=0$, $V_a=V_b=-\tilde{Y}$/2 and $t_{ab}=\tilde{X}$. Energy levels obtained numerically with this method have a good matching shown in Fig. \ref{fig:hbh4}. 
 Note that if an interspecies nearest neighbor interaction $V_{ab}$ is introduced, it can be incorporated in $\mathcal{H}_{eff}$ by making the substitution $V_a+V_b\rightarrow V_a+V_b-2V_{ab}$ in the first term of Eq. (\ref{eq:H_eff}). 
 
 This is a very different realization than for the $O(2)$ limit. The presence in the Hamiltonian of the additional term that interchanges the species index, Eq. (37), rules out implementing this Hamiltonian using mixtures of two different types of atoms, and it also makes it difficult to realize it using two hyperfine states of the same atomic species. It could be realized with a single atomic species on a ladder structure with $a$ and $b$ corresponding to the two legs of the ladder. Ladder systems have been realized experimentally by using lattices of double wells~\cite{2006PhRvA..73c3605S,2007Natur.448.1029F,2011NatPh...7...61C,2013Sci...340.1307G,2014NatPh..10..588A}.
The hopping amplitudes can be tuned such that the hopping in the direction along the ladder is negligible, but finite along the rungs, thus exchanging $a$, $b$ species index at the same rung.
An attractive intraspecies interaction ($V_a=V_b=-\tilde{Y}/2$) is also needed, favoring having two atoms in neighboring sides on the same leg of the ladder. This acts as a nearest-neighbor ferromagnetic coupling in the effective one-dimensional spin chain Hamiltonian. For the experimental implementation, an attractive nearest neighbor interaction can be obtained by using cold dipolar atoms or molecules, with dipole moments aligned along the ladder and with inter-rung distance such that the rapidly decaying dipole-dipole interaction between next-nearest-neighbors can be neglected. With this alignment of the moments, the inter-species interaction in the same rung ($W$) is repulsive.
Experiments with ultra cold dipolar quantum gases have been performed with chromium \cite{2005PhRvL..94p0401G}, erbium \cite{2012arXiv1204.1725A,2014PhRvL.112a0404A}, and dysprosium \cite{2011PhRvL.107s0401L,2012PhRvL.108u5301L,2014arXiv1411.3069T}, which have magnetic moments $6 \mu _B$, $7 \mu _B$ and $10 \mu _B$, respectively, and with polar molecules, such as $^{40}$K $^{87}$Rb and $^{23}$Na $^{40}$K~\cite{2008Sci...322..231N,2012PhRvL.108h0405C,2012PhRvL.109h5301W}.

\section{Conclusions}
With the remarkable experimental progress in the field of Atomic, Molecular and Optical physics, particularly the loading of trapped ultra-cold atoms onto optical lattices, and the unprecedented levels of control and tunability of their interactions, a wide range of quantum many-body lattice Hamiltonians can be realized and studied.

We have proposed a two-species Bose-Hubbard model that may be used as a quantum  simulator for lattice gauge theories. The gauge invariance is built-in and thus does not need to be achieved via fine-tuning, and the correspondence between the proposed Bose-Hubbard Hamiltonian and the Abelian Higgs model can be checked quantitatively.

On the experimental side, various multi-species Bose-Hubbard systems can be created with current cold atom technology. Various possible realizations for the two species that are needed can be explored. On the theory side, the Bose-Hubbard Hamiltonian is amenable to Quantum Monte Carlo calculations and the Abelian Higgs model can be treated with the TRG (and the time-continuum limit obtained) as well as Monte Carlo calculations. The full spectra can be calculated, at least for small systems. The TRG formulation can very naturally provide time-dependent correlation functions. Therefore, beside the ultimate goal of quantum simulating lattice gauge theories using cold atoms on optical lattices, the correspondence we report here may also provide a way of studying the dynamics of the cold atom system such as response to sudden quenches.

Acknowledgments. We thank  N. Gemelke, E. Mueller, Li-Ping Yang, J. Osborn, and Yuzhi Liu for stimulating discussions. This article was completed while Y. M. was attending the workshop 
INT Program INT-15-1 ``Frontiers in Quantum Simulation with Cold Atoms", he thanks the participants for many interesting discussions. 
This research was supported in part  by the Department of Energy
under Award Numbers DOE grant DE-FG02-05ER41368, DE-SC0010114 and DE-FG02-91ER40664, the NSF under grant DMR-1411345 and by the Army Research Office of the Department of Defense under Award Number W911NF-13-1-0119.  

\begin{thebibliography}{10}
\bibitem{nextpub}%
  \BibitemOpen
  \bibinfo {note} {The details will be presented in a separate publication}%
  \bibAnnoteFile{NoStop}{nextpub}%
\end{thebibliography}
\end{document}